\documentclass{webofc}

\newcommand {\jpsi}      {\ensuremath{\mathrm{J}\kern-0.02em/\kern-0.05em\psi} }
\newcommand {\pt}        {\ensuremath{p_{\mathrm{T}}}}
\newcommand {\y}         {\ensuremath{y} }
\newcommand {\pp}        {\ensuremath{\mathrm {p\kern-0.05em p}}}
\newcommand {\PbPb}      {\ensuremath{\mathrm{Pb\mbox{--}Pb}}~}
\newcommand {\gev}       {\ensuremath{\,\mathrm{GeV}}}

\newcommand {\gevc}      {\ensuremath{\,\mathrm{GeV}\kern-0.05em/\kern-0.02em c}}
\newcommand{\RAA}{$R_{\mathrm{AA}}$ }

\usepackage[varg]{txfonts}   
\usepackage{hyperref}
\usepackage{url}
\usepackage{lineno}
\hypersetup{colorlinks=true,citecolor=blue,urlcolor=blue,linkcolor=blue}
\begin{document}
\title{Investigating the interplay between initial hard processes and final-state effects measuring prompt and non-prompt \jpsi}
%
%

\author{\firstname{Maurice} \lastname{Coquet}\inst{1}\fnsep\thanks{\email{maurice.louis.coquet@cern.ch}}, 
        on behalf of the ALICE Collaboration
}

\institute{SUBATECH, Nantes University, IMT Atlantique, IN2P3/CNRS, 4 rue Alfred Kastler, 44307 Nantes, France}
\modulolinenumbers[1]
\abstract{Charmonium production in high-energy collisions can be split into a prompt and a non-prompt component. Both components can be distinguished experimentally by studying displaced topology, and represent valuable tools to investigate the properties of the strongly interacting medium produced in ultra-relativistic heavy-ion collisions. In particular, the study of non-prompt charmonium production can give access to the beauty-hadron production cross section and can be used to investigate the in-medium energy loss of beauty quarks. In these proceedings, the recent measurement of prompt and non-prompt \jpsi carried out by the ALICE Collaboration in \PbPb collisions at midrapidity ($|\y|$ < 0.8) at $\sqrt{s_{\rm NN}}=5.02$ TeV, will be presented. Moreover, thanks to the installation of the new muon forward tracker (MFT), the prompt/non-prompt charmonium separation will be possible in LHC Run 3 also at forward rapidity (2.5 < \y < 4).}
\maketitle
\section{Introduction}
\label{intro}
Heavy quarks, charm and beauty quarks, are excellent probes to study the strongly interacting medium produced in ultra-relativistic heavy-ion collisions, the quark-gluon plasma (QGP). Indeed, due to their large masses, these quarks are produced by the initial hard scatterings at the beginning of the collision and experience the full evolution of the collision. In particular, charmonia, bound states of charm and anti-charm quarks, are extensively studied in heavy-ion collisions to access properties of the QGP and its evolution. In such collisions, the production of charmonium is affected by mechanisms which modify their production with respect to smaller collision systems. First, charmonium states are expected to experience dissociation in the QGP by color-charge screening and dynamical dissociation through interactions with medium constituents~\cite{Rothkopf:2019ipj}. In addition, at LHC energies, a large number of $c\Bar{c}$ pairs is produced, sufficient so that charmonium production is enhanced via the recombination of uncorrelated $c\Bar{c}$ pairs~\cite{Braun-Munzinger:2000csl, Thews:2000rj}. This mechanism is known as regeneration.
Finally, gluon energy loss in the QGP is expected to affect the production of heavy quarks, and thus of charmonium, at high \pt~\cite{Arleo:2017ntr}, and cold nuclear matter effects such as nuclear shadowing of the parton distribution functions~\cite{Apolinario:2022vzg} must also be taken into account in charmonium production.
A commonly studied charmonium state is the vector state \jpsi. Its production in high-energy collisions includes prompt and non-prompt contributions, which can be distinguished based on the displacement of the production vertex of the state. On the one hand, prompt \jpsi are produced directly at the collision vertex between the two colliding hadrons, or from the feed-down of excited charmonium states. Hence, prompt \jpsi can probe the suppression and regeneration mechanisms inside the QGP. On the other hand, non-prompt \jpsi are produced through the decay of hadrons containing beauty quarks. Their study can thus provide information on the partonic energy loss of the beauty quark, as well as its transport coefficients inside the QGP.

With the ALICE apparatus, charmonium can be measured either at midrapidity ($|\y|$ < 0.8) in
the dielectron decay channel, or at forward rapidity (2.5 < \y < 4) in the dimuon decay channel. In Run 2, the separation between prompt and non-prompt contribution to \jpsi production was only achievable at midrapidity, using the Inner Tracking System (ITS) for vertexing, while the forward muon arm was able to perform only inclusive measurements. A detailed description of the ALICE apparatus can be found in Ref.\cite{ALICE:2008ngc}.

\section{Results}
\label{results}

In Run 2, the fraction of non-prompt \jpsi was measured by ALICE at midrapidity in \PbPb collisions at 5.02 TeV, down to \pt~= 1.5 \gevc~\cite{ALICE:2023hou}. The extraction of this fraction is based on a two-dimensional maximum-log-likelihood fit using two variables: the invariant mass of the dielectron candidates, and the pseudo-proper decay length $x$ defined as $x=c \times L_{\mathrm{xy}} \times m_{\mathrm{J} / \psi} / p_{\mathrm{T}}$, where $L_{\mathrm{xy}}=(\vec{L} \times\vec{p_{\rm T}}) / p_{\mathrm{T}}$ is the projection in the transverse plane of the vector pointing from the primary vertex to the \jpsi decay vertex, and $m_{\mathrm{J} / \psi}$ is the \jpsi mass. 

\begin{figure}[h]
    \centering
    \rotatebox{0}{
    \includegraphics[width=0.475\textwidth]{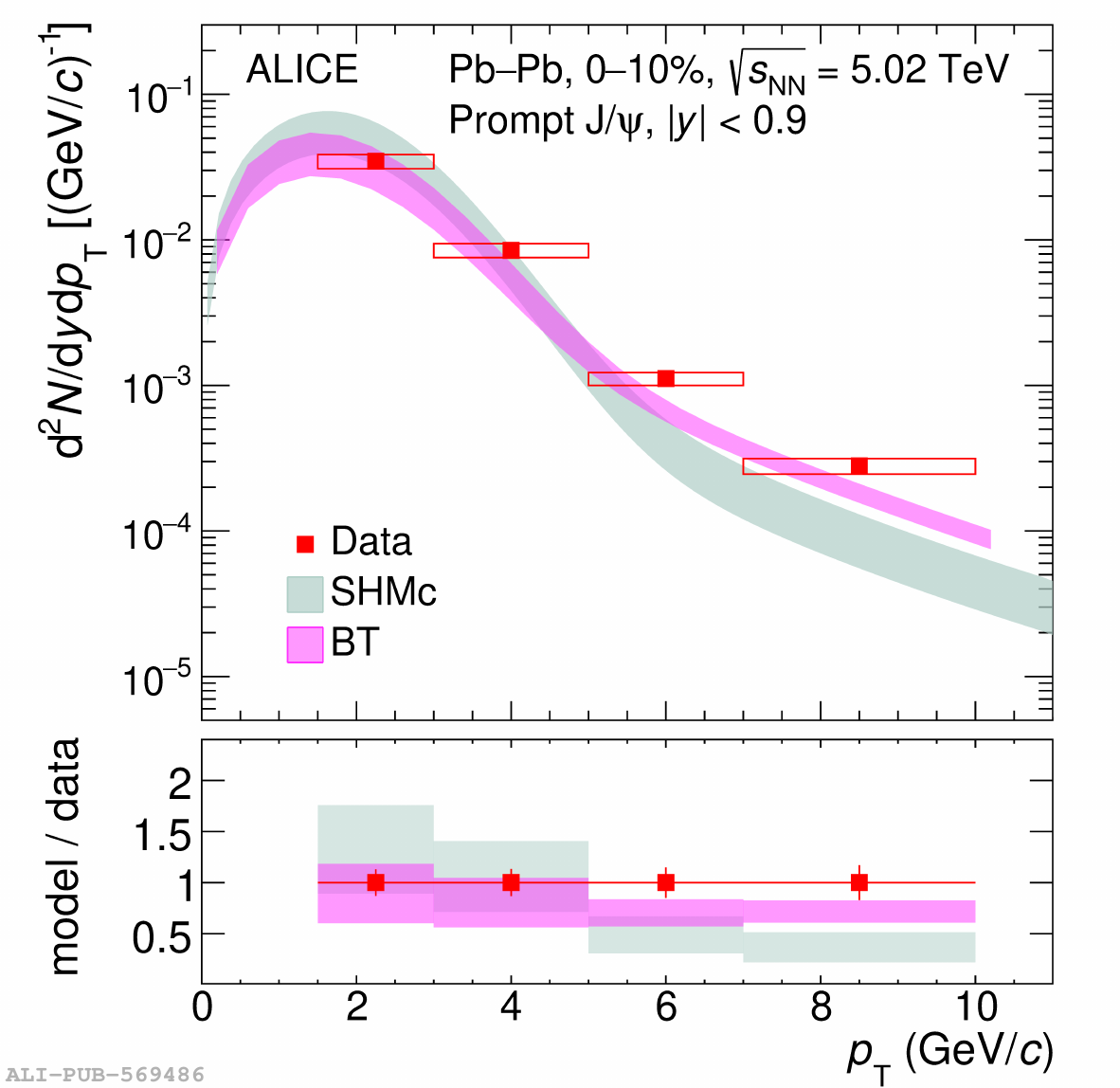}
    \includegraphics[width=0.475\textwidth]{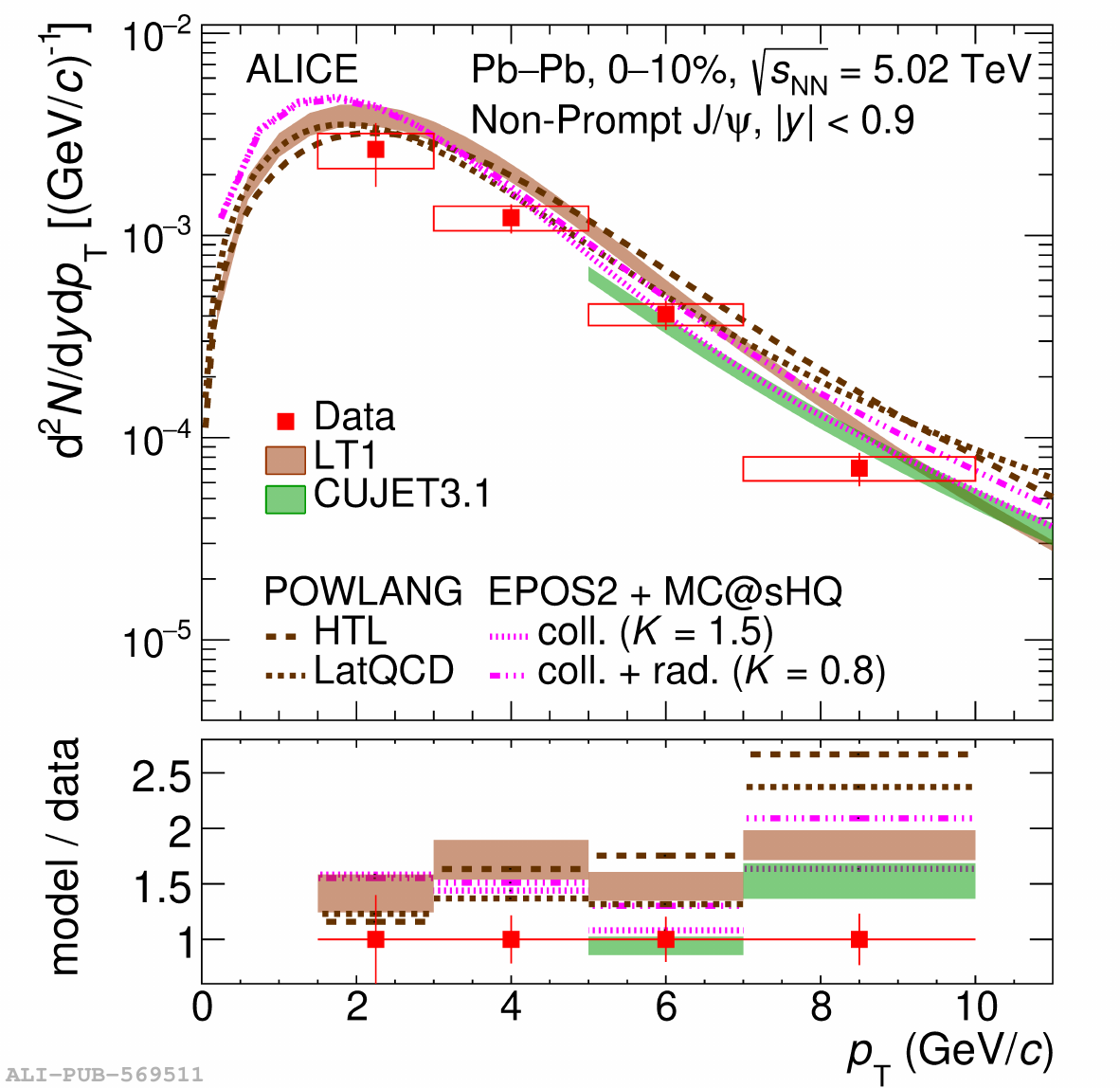}
    }
    \caption{Prompt (left) and non-prompt (right) \jpsi yields as a function of \pt~in the $0-10 \%$ centrality class compared with models~\cite{ALICE:2023hou}. Vertical error bars and boxes represent statistical and systematic uncertainties, respectively. Bottom panels show the ratios between models and data.}
    \label{fig:cross-section}
\end{figure}
The prompt and non-prompt \jpsi \pt~spectra in \PbPb are computed by combining the non-prompt fraction with the measured inclusive \jpsi \pt-dependent cross section~\cite{ALICE:2023gco}. The obtained cross sections are compared with available models. The statistical hadronization model (SHMc)~\cite{Andronic:2019wva} assumes that all bound states are produced at hadronization and that their yields follow thermal weights. The Boltzmann transport model (BT) is based on a detailed balance equation, involving continuous dissociation and regeneration of \jpsi states during the QGP phase~\cite{Chen:2018kfo}. Figure~\ref{fig:cross-section} (left) shows a comparison of these two models with the ALICE \PbPb data in the 0-10\% centrality class. For \pt~< 5 \gevc, both models are in agreement with the data, while for \pt~> 5 \gevc, the statistical model tends to underpredict the data. In SHMc, two contributions to \jpsi production are considered: core, which corresponds to thermal production at phase boundary, and corona, corresponding to the cooler edges of the fireball where no QGP is expected to form, and where \jpsi production is assumed to be captured by scaled pp distributions. The uncertainty associated with SHMc plotted on Fig.\ref{fig:cross-section} (left) corresponds to variation of the fraction of corona \jpsi. For the BT model, the uncertainty corresponds to variation of the initial $c\Bar{c}$ cross section.

The non-prompt cross section is compared on Fig.\ref{fig:cross-section} (right) with models based on Langevin equations, such as LT1~\cite{Yang:2023rgb} or POWLANG~\cite{Beraudo:2021ont}, as well as jet energy-loss framework (CUJET~\cite{Shi:2018izg}) and EPOS2+MC@sHQ~\cite{Nahrgang:2016lst} which employs the EPOS2 event generator coupled with a Monte Carlo treatment of Boltzmann transport for the heavy quarks. As shown on Fig.\ref{fig:cross-section} (right), these models overpredict the data for \pt~> 7 \gevc. This is particularly the case for the POWLANG model which includes only collisional energy loss for the b-quark, while no radiative contribution to the energy loss is taken into account. This model provides two different predictions, corresponding to two evaluations of the b quark transport coefficients, one using Hard Thermal Loop perturbative calculations (HTL) and the other one computed from lattice QCD (LatQCD). The uncertainty associated with the LT1 model shown on Fig.\ref{fig:cross-section} (right) corresponds to variations of the nuclear shadowing taken into account in the initial heavy-quark production. The CUJET uncertainties correspond to variations of the running QCD coupling and of the magnetic screening mass.

\begin{figure}[h]
    \centering
    \rotatebox{0}{
    \includegraphics[width=0.475\textwidth]{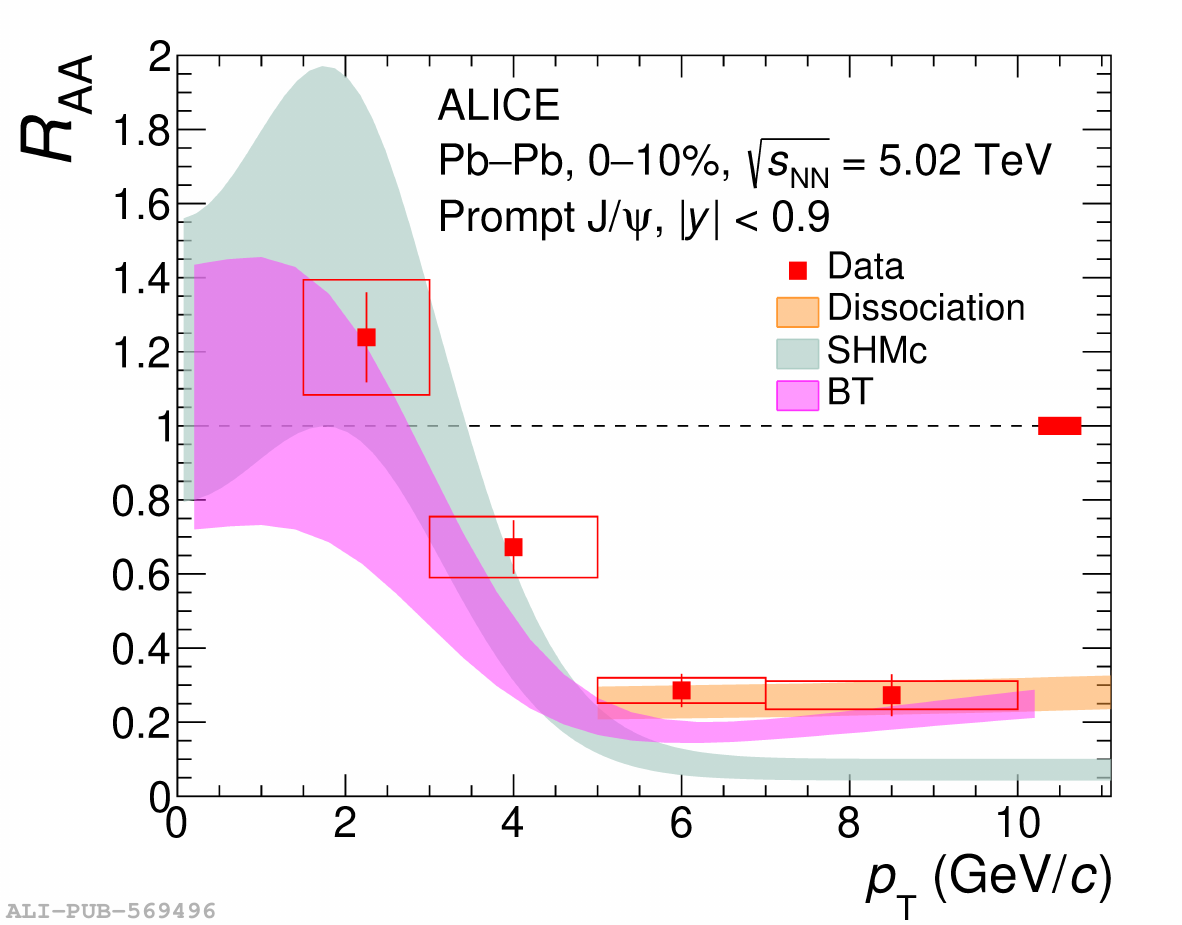}
    \includegraphics[width=0.475\textwidth]{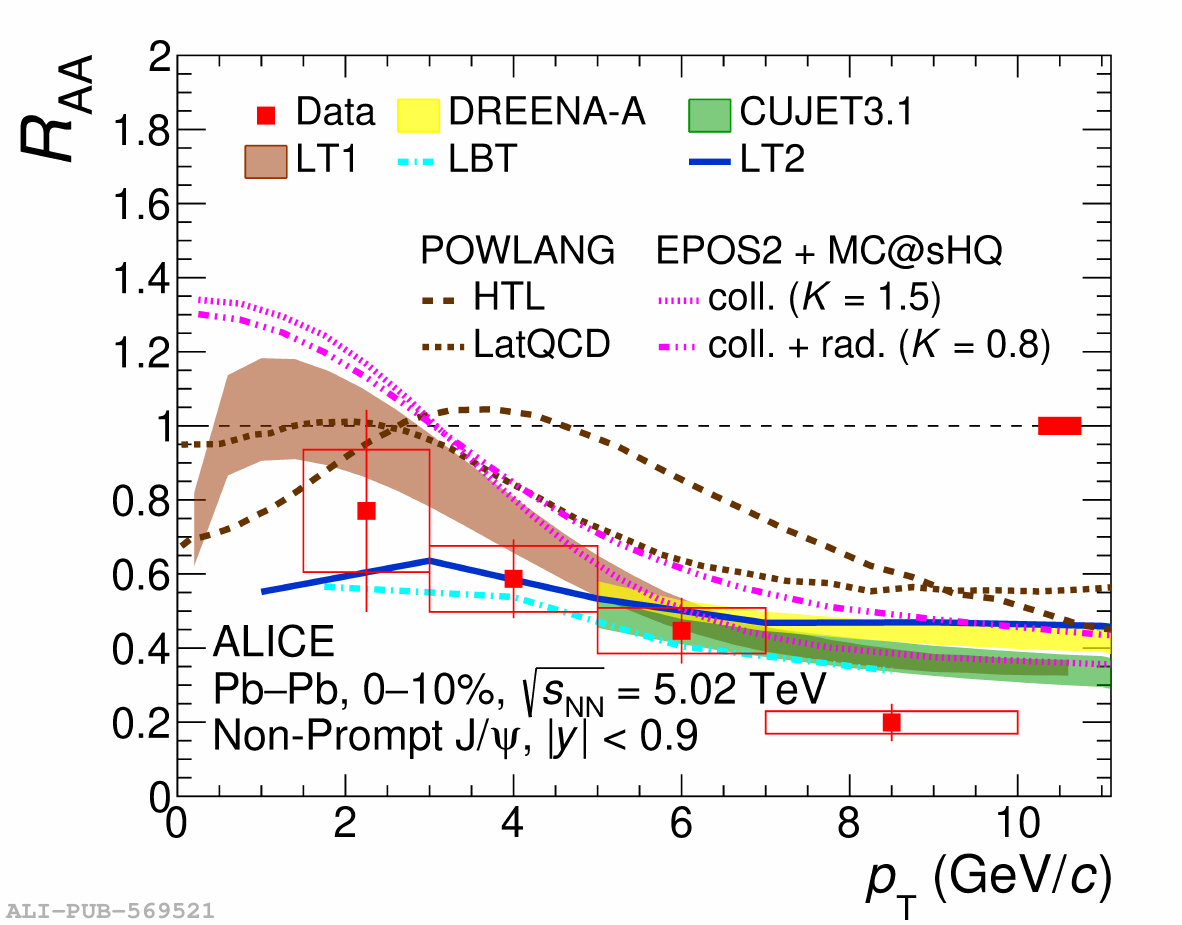}
    }
    \caption{Prompt (left) and non-prompt \jpsi \RAA as a function of \pt~in the $0-10 \%$ centrality class, compared with models~\cite{ALICE:2023hou}. Error bars and boxes represent statistical and uncorrelated systematic uncertainties, respectively.}
    \label{fig:RAA}
\end{figure}

Using the measured cross sections in \PbPb and in pp, the nuclear modification factor \RAA can be computed. 
Figure \ref{fig:RAA} (left) shows comparisons of the prompt \RAA with models predictions. For \pt~> 5 \gevc, the BT model lies on the lower side of the data, while the SHMc model underestimates the data. Both models correctly describe the measured \RAA for \pt~< 5 \gevc. A dissociation model is also shown~\cite{Aronson:2017ymv}, based on a rate equation using a decay rate computed from momentum broadening of the charm quarks. It is in good agreement with the data at high \pt. The uncertainty on the predictions from the dissociation model shown on Fig.\ref{fig:RAA} (left) corresponds to variations of the formation time of the charmonium state. 

The non-prompt \RAA is shown of Fig.\ref{fig:RAA} (right) and compared with models. The POWLANG model overpredicts the data, for both estimates of the transport coefficients. The EPOS2+MC@sHQ model tends to overpredict the data for \pt~< 3 \gevc, while it best encapsulates the high-\pt~region by including only collisional energy loss. Additional models are also shown. The LT2 model~\cite{Li:2021xbd} is another Langevin transport approach.
The LBT model~\cite{Xing:2021xwc} is based on linear Boltzmann transport, which includes both short and long-range confining interactions in the heavy-quark potential. LT1, LT2 and LBT all include both radiative and collisional energy loss, and are in agreement with the data over the whole \pt~range. Finally, DREENA~\cite{Zigic:2021rku} is based on a state-of-the-art energy-loss framework coupled with hydrodynamic description of the medium and provides predictions for \pt~> 5 \gevc. DREENA and CUJET3.1 both overpredict the data for \pt~> 7 \gevc.

\section{Summary and outlook}
\label{sum}

New ALICE results were shown for prompt/non-prompt \jpsi separation in \PbPb collisions down to \pt~= 1.5 \gevc. The prompt data are described by models including regeneration of uncorrelated $c\Bar{c}$ pairs at low \pt~and charmonium suppression through color screening and dynamical dissociation at high \pt. The non-prompt \jpsi data are described by several models. For \pt~> 5 \gev, all calculations are consistent with the data except for POWLANG, while at low \pt, EPOS2+MC@sHQ overpredits the data. The models which are in best agreement over the full \pt~range are the ones which include both collisional and radiative energy loss mechanisms for the beauty quark.


For Run 3, ALICE underwent several important upgrades~\cite{ALICE:2023udb}. In particular, new detectors were installed, such as the Muon Forward Tracker (MFT). This new detector allows to perform precise vertexing for the muon spectrometer, and, as a consequence, to separate prompt and non-prompt \jpsi at forward rapidity in the dimuon decay channel. A first evaluation of the performance for prompt/non-prompt separation at forward rapidity was carried out in \pp~collisions. It showed that this separation is achievable down to \pt~= 0, thanks to the longitudinal boost of particles detected in this region, and opens the door to future prompt/non-prompt \jpsi analysis in \PbPb at forward rapidities, a phase-space region yet unexplored at LHC.

%
%
%

\end{document}